\documentclass[12pt]{article}
\usepackage{amsmath,amssymb}
\usepackage{bm}
\newcommand{\eqdef}{\stackrel{\text{def}}{=}}
\newcommand{\n}{\nonumber \\}
\newcommand{\ignore}[1]{}

\allowdisplaybreaks[3]

\begin{document}

\baselineskip=20pt

%%%%%%%%%%%%%%%%%%%%%%%%%%%%%%%%%%%%%%%%%%%%%%%%%%%%%%%%%%%%
%                                                          %
%  Title page                                              %
%                                                          %
%%%%%%%%%%%%%%%%%%%%%%%%%%%%%%%%%%%%%%%%%%%%%%%%%%%%%%%%%%%%
\newfont{\elevenmib}{cmmib10 scaled\magstep1}
\newcommand{\preprint}{
    \begin{flushleft}
      \elevenmib Yukawa\, Institute\, Kyoto\\
    \end{flushleft}\vspace{-1.3cm}
    \begin{flushright}\normalsize  \sf
      DPSU-07-2\\
      YITP-07-39\\
%      {\tt arXiv:YYMM.NNNN}\\ %% not replaced automatically!!
      June 2007
    \end{flushright}}
\newcommand{\Title}[1]{{\baselineskip=26pt
    \begin{center} \Large \bf #1 \\ \ \\ \end{center}}}
\newcommand{\Author}{\begin{center}
    \large \bf Satoru~Odake${}^a$, Yama\c{c}~Pehlivan${}^b$
and Ryu~Sasaki${}^c$ \end{center}}
\newcommand{\Address}{\begin{center}
      $^a$ Department of Physics, Shinshu University,\\
      Matsumoto 390-8621, Japan\\
      ${}^b$ Department of Physics, University of Wisconsin-Madison,\\
      1150 University Avenue, Madison WI 53706, USA\\
      ${}^c$ Yukawa Institute for Theoretical Physics,\\
      Kyoto University, Kyoto 606-8502, Japan
    \end{center}}
\newcommand{\Accepted}[1]{\begin{center}
    {\large \sf #1}\\ \vspace{1mm}{\small \sf Accepted for Publication}
    \end{center}}

\preprint \thispagestyle{empty}
\bigskip\bigskip\bigskip

\Title{Interpolation of SUSY quantum mechanics} \Author

\Address

\begin{abstract}
Interpolation of two adjacent Hamiltonians in SUSY quantum
mechanics $\mathcal{H}_s\eqdef(1-s)\mathcal{A}^\dagger\mathcal{A}
+s\mathcal{A}\mathcal{A}^\dagger$, $0\le s\le1$ is discussed
together with related operators. For a wide variety of
shape-invariant degree one quantum mechanics and their `discrete'
counterparts, the interpolation Hamiltonian is also
shape-invariant, that is it takes the same form as the original
Hamiltonian with shifted coupling constant(s).
\end{abstract}

%%%%%%%%%%%%%%%%%%%%%%%%%%%%%%%%%%%%%%%%%%%%%%%%%%%%%%%%%%%%%%%
%                                                             %
%  1. Introduction                                            %
%                                                             %
%%%%%%%%%%%%%%%%%%%%%%%%%%%%%%%%%%%%%%%%%%%%%%%%%%%%%%%%%%%%%%%
\section{Introduction}
\label{intro} \setcounter{equation}{0}

The factorisation method \cite{crum} or the so-called
super-symmetric (SUSY) quantum mechanics \cite{ew,susyqm} is a
well-established tool for investigating degree one quantum
mechanics including their `discrete' counterparts \cite{os4os5}.
In its essence the SUSY quantum mechanics asserts that a
factorised Hamiltonian
$\mathcal{H}\eqdef\mathcal{A}^\dagger\mathcal{A}$ and its reversed
order (SUSY-partner) Hamiltonian
$\mathcal{H}_r\eqdef\mathcal{A}\mathcal{A}^\dagger$ are {\em
iso-spectral\/} except for the ground state. Let us denote by
$\phi_n$ the eigenfunction of $\mathcal{H}$:
\begin{equation}
  \mathcal{H}\phi_n=\mathcal{E}_n\phi_n,\quad n=0,1,2,\ldots\quad
 \mathcal{E}_0<\mathcal{E}_1<\cdots .
\end{equation}
Then $\tilde{\phi}_n\eqdef \mathcal{A}\phi_n$ is also an
eigenfunction of $\mathcal{H}_r$ with the same eigenvalue
\begin{equation}
  \mathcal{H}_r\tilde{\phi}_n=\mathcal{E}_n\tilde{\phi}_n,\quad
  n=1,2,\ldots,
\end{equation}
except for $\phi_0$, which is annihilated by $\mathcal{A}$
\begin{equation}
  \mathcal{A}\phi_0=0\quad\bigl(\Rightarrow
  \mathcal{H}\phi_0=0,\quad \mathcal{E}_0=0\bigr).
\end{equation}

In various contexts of quantum physics, one encounters quite often
a situation \cite{balpeh} in which an interpolation of the two
super-symmetric partner Hamiltonians
\begin{equation}
  \mathcal{H}_s\eqdef(1-s)\mathcal{A}^\dagger\mathcal{A}
  +s\mathcal{A}\mathcal{A}^\dagger, \quad 0\le s\le1,
  \label{intham}
\end{equation}
or an operator closely related with it, plays an important role.
For a wide class of {\em shape-invariant\/} Hamiltonians
\cite{os4os5,genden,os6}, we show in this paper that the
interpolation also {\em retains shape-invariance\/}. That is the
interpolating Hamiltonian \eqref{intham} has the same form as the
original Hamiltonian with shifted coupling constant(s) and a
shifted ground state energy.

This paper is organised as follows. In section two the basic facts
and notation of shape-invariant quantum mechanics are
recapitulated. In section three the assertion of the
shape-invariant interpolation is demonstrated for various examples
of shape-invariant potentials in ordinary quantum mechanical
systems \cite{susyqm,genden,os6}. They have the classical
orthogonal polynomials, the Hermite, Laguerre and Jacobi
polynomials as a part of the eigenfunctions. In section four, we
demonstrate the assertion, in a slightly different form, for
various `discrete' quantum mechanical systems \cite{os4os5,os6}
which have the Askey-Wilson, Wilson, continuous dual Hahn,
continuous Hahn and Meixner-Pollaczek polynomials
\cite{And-Ask-Roy,koeswart} as a part of the eigenfunctions. These
polynomials belong to the Askey scheme of hypergeometric
orthogonal polynomials and they are the deformations of the
Jacobi, Laguerre and Hermite polynomials. In section five we apply
the method presented in the previous section to the ordinary
quantum mechanical systems. This gives another type of
interpolation of the ordinary SUSY quantum mechanics. The final
section is for a summary and comments.

%%%%%%%%%%%%%%%%%%%%%%%%%%%%%%%%%%%%%%%%%%%%%%%%%%%%%%%%%%%%%%%
%                                                             %
%  2. Shape-Invariant Quantum Mechanics                       %
%                                                             %
%%%%%%%%%%%%%%%%%%%%%%%%%%%%%%%%%%%%%%%%%%%%%%%%%%%%%%%%%%%%%%%
\section{Shape-Invariant Quantum Mechanics}
\label{shape} \setcounter{equation}{0}

Here we discuss only the degree one quantum mechanics. A
shape-invariant \cite{genden} quantum mechanical system consists
of a series of isospectral (and factorised) Hamiltonians
$\{\mathcal{H}(\bm{\lambda})\}$ parametrised by (a set of)
parameters $\bm{\lambda}=(\lambda_1,\lambda_2,\cdots)$:
\begin{equation}
  \mathcal{H}(\bm{\lambda})\eqdef\mathcal{A}(\bm{\lambda})^{\dagger}
  \mathcal{A}(\bm{\lambda}),\quad
  \mathcal{H}(\bm{\lambda}) \phi_n(x\,;\bm{\lambda})
  =\mathcal{E}_n(\bm{\lambda})\phi_n(x\,;\bm{\lambda}),\quad
  n=0,1,2,\ldots.
\end{equation}
Shape invariance simply means that the reversed order Hamiltonian
\begin{equation}
  \mathcal{H}_r(\bm{\lambda})\eqdef\mathcal{A}(\bm{\lambda})
  \mathcal{A}(\bm{\lambda})^{\dagger},
  \label{Hr}
\end{equation}
takes the same form (shape) as the original Hamiltonian with
`shifted' parameters:
\begin{equation}
  \mathcal{A}(\bm{\lambda})\mathcal{A}(\bm{\lambda})^{\dagger}
  =\mathcal{A}(\bm{\lambda}+\bm{\delta})^{\dagger}
  \mathcal{A}(\bm{\lambda}+\bm{\delta})+\mathcal{E}_1(\bm{\lambda}),
  \label{shapeinv}
\end{equation}
where $\bm{\delta}$ is the shift of the parameter and
$\mathcal{E}_1(\bm{\lambda})$ is the increase of the ground state
energy. Note  that the entire energy spectrum is determined by
shape-invariance, since
$\mathcal{E}_{n+1}(\bm{\lambda})=\mathcal{E}_n(\bm{\lambda})
+\mathcal{E}_1(\bm{\lambda}+n\bm{\delta})$, which implies
$\{\mathcal{E}_n\}$ are determined in terms of $\mathcal{E}_1$,
$\mathcal{E}_n(\bm{\lambda})
=\sum_{k=0}^{n-1}\mathcal{E}_1(\bm{\lambda}+k\bm{\delta})$
\cite{os4os5,os6}.

With this notation, our assertion goes as follows:
\begin{equation}
  \mathcal{H}_s\eqdef(1-s)\mathcal{H}+s \mathcal{H}_r
  =\mathcal{H}(\bm{\lambda}')+\Delta\mathcal{E}(\bm{\lambda},s),
  \label{mainresult}
\end{equation}
in which the new coupling constants $\bm{\lambda}'$ satisfy the
boundary condition
\begin{eqnarray}
  \text{for}\ s=0:\ \bm{\lambda}'=\bm{\lambda},\quad
  \text{and for}\ s=1:\ \bm{\lambda}'=\bm{\lambda}+\bm{\delta}.
\end{eqnarray}
The constant part $\Delta\mathcal{E}(\bm{\lambda},s)$ is the shift
of the ground state energy. It also satisfies the boundary
condition
\begin{eqnarray}
  \text{for}\ s=0:\ \Delta\mathcal{E}(\bm{\lambda},0)=0,\quad
  \text{and for}\ s=1:\ \Delta\mathcal{E}(\bm{\lambda},1)
  =\mathcal{E}_1(\bm{\lambda}).
\end{eqnarray}
Here $\mathcal{E}_1(\bm{\lambda})$ is the last term in
\eqref{shapeinv}.

If our assertion \eqref{mainresult} holds, the eigenfunctions and
eigenvalues of $\mathcal{H}_s$ are given by
\begin{equation}
  \mathcal{H}_s\phi_n(x\,;\bm{\lambda}')
  =\bigl(\mathcal{E}_n(\bm{\lambda}')+\Delta\mathcal{E}(\bm{\lambda},s)
  \bigr)\phi_n(x\,;\bm{\lambda}'),\quad n=0,1,2,\ldots.
\end{equation}
Note that $\bm{\lambda}'$ depend on $s$.

%%%%%%%%%%%%%%%%%%%%%%%%%%%%%%%%%%%%%%%%%%%%%%%%%%%%%%%%%%%%%%%
%                                                             %
%  3. Ordinary Quantum Mechanics                              %
%                                                             %
%%%%%%%%%%%%%%%%%%%%%%%%%%%%%%%%%%%%%%%%%%%%%%%%%%%%%%%%%%%%%%%
\section{Ordinary Quantum Mechanics}
\label{ordinaryqm} \setcounter{equation}{0}

The ordinary quantum mechanics is quite conveniently described by
the {\em prepotential} $W(x\,;\bm{\lambda})$ $\in\mathbb{R}$ which
parametrises the ground state wavefunction
\begin{equation}
  \phi_0(x\,;\bm{\lambda})\propto e^{W(x\,;\bm{\lambda})}.
\end{equation}
This is possible because the ground state wavefunction does not
have nodes and can be chosen real. Then the factorised Hamiltonian
reads
\begin{align}
  &\mathcal{H}=\mathcal{A}(\bm{\lambda})^\dagger\mathcal{A}(\bm{\lambda})
  =-\frac{d^2}{dx^2}+\Bigl(\frac{dW(x\,;\bm{\lambda})}{dx}\Bigr)^2
  +\frac{d^2W(x\,;\bm{\lambda})}{dx^2},\\
  &\mathcal{A}(\bm{\lambda})\eqdef
  -\frac{d}{dx}+\frac{dW(x\,;\bm{\lambda})}{dx},\quad
  \mathcal{A}(\bm{\lambda})^\dagger=
  \frac{d}{dx}+\frac{dW(x\,;\bm{\lambda})}{dx}.
\end{align}
Here we have adopted the unit system $m=\hbar=1$ and the overall
normalisation $\mathcal{H}=-\frac{d^2}{dx^2}+\cdots$. Obviously
the ground state is annihilated by the operator $\mathcal{A}$
\begin{equation}
  \mathcal{A}(\bm{\lambda})\phi_0(x\,;\bm{\lambda})=0\quad
  \bigl(\Rightarrow\mathcal{H}\phi_0(x\,;\bm{\lambda})=0,\quad
  \mathcal{E}_0(\bm{\lambda})=0\bigr).
\end{equation}
We assume the shape-invariance \eqref{shapeinv}. In this form the
interpolating Hamiltonian $\mathcal{H}_s$ reads simply
\begin{equation}
  \mathcal{H}_s=-\frac{d^2}{dx^2}
  +\Bigl(\frac{dW(x\,;\bm{\lambda})}{dx}\Bigr)^2
  +(1-2s)\frac{d^2W(x\,;\bm{\lambda})}{dx^2}
  =\mathcal{H}-2s\,\frac{d^2W(x\,;\bm{\lambda})}{dx^2}.
\end{equation}

\medskip
Let us consider the known typical examples of shape-invariant
potentials. All of them satisfy our assertion \eqref{mainresult}.

%%%%%%%%%%%%%%%%%%%%%%%%%%%%%%%%%%%%%%%%%%%%%%%%%
%                                               %
%  3.1. harmonic oscillator: a trivial example  %
%                                               %
%%%%%%%%%%%%%%%%%%%%%%%%%%%%%%%%%%%%%%%%%%%%%%%%%
\subsection{harmonic oscillator: a trivial example}
\label{ho}

The harmonic oscillator is a trivial example with
$W(x\,;\omega)=-\omega x^2/2$. This simply leads to the same
Hamiltonian with the shifted ground state energy
\begin{equation}
  \mathcal{H}_s=\mathcal{H}+\Delta\mathcal{E}(\omega,s),\qquad
  \Delta\mathcal{E}(\omega,s)=2s\omega.
\end{equation}

%%%%%%%%%%%%%%%%%%%%%%%%%%%%%%%%%%%%%%%%%%%%%%%%%
%                                               %
%  3.2. harmonic oscillator with a centrifugal  %
%       barrier potential                       %
%                                               %
%%%%%%%%%%%%%%%%%%%%%%%%%%%%%%%%%%%%%%%%%%%%%%%%%
\subsection{harmonic oscillator with a centrifugal barrier potential}
\label{ho_cbp}

This is the one-body case of the Calogero model \cite{cal} with
\begin{align}
  &\bm{\lambda}=(\omega,g),\quad \bm{\delta}=(0,1),\quad
  \mathcal{E}_n(\omega,g)=4n, \quad n=0,1,2,\ldots,\\
  &W(x\,;\omega,g)=-\omega x^2/2+g\log x,\quad \omega>0,\quad g\geq 1/2,
  \quad x>0,\\
  &\mathcal{A}(\omega,g)=-\frac{d}{dx}-\omega x+\frac{g}{x},\quad
  \mathcal{A}(\omega,g)^{\dagger}=\frac{d}{dx}-\omega x+\frac{g}{x}.
\end{align}
The original and the interpolating Hamiltonians read
\begin{align}
  \mathcal{H}(\omega,g)&=-\frac{d^2}{dx^2}+\omega^2x^2
  +\frac{g(g-1)}{x^2}-\omega(1+2g),
  \label{oldcalham}\\
  \mathcal{H}_s&=-\frac{d^2}{dx^2}+\omega^2x^2+\frac{g(g+2s-1)}{x^2}
  -\omega(1+2g-2s).
  \label{s-calham}
\end{align}
It is obvious that the interpolating Hamiltonian $\mathcal{H}_s$
has the same form as the old \eqref{oldcalham} with the coupling
constant $g$ replaced by $g'$:
\begin{equation}
  g'=\bigl(1+\sqrt{1+4g(g+2s-1)}\,\bigr)/2
  \label{gprimeeq}
\end{equation}
together with the shift of the ground state energy
$\Delta\mathcal{E}(\omega,g,s)=2\omega(s+g'-g)$.

%%%%%%%%%%%%%%%%%%%%%%%%%%%%%%%%%%%%%%%%%%%%%%%%%
%                                               %
%  3.3. $1/\sin^2x$ potential or symmetric      %
%       P\"oschl-Teller potential               %
%                                               %
%%%%%%%%%%%%%%%%%%%%%%%%%%%%%%%%%%%%%%%%%%%%%%%%%
\subsection{$1/\sin^2x$ potential or symmetric P\"oschl-Teller potential}
\label{sPT}

This is the one-body case of the well-known Sutherland model
\cite{sut}. The prepotential, the Hamiltonian and other data are:
\begin{align}
  &\bm{\lambda}=g,\quad \bm{\delta}=1,\quad
  \mathcal{E}_n(g)=n(n+2g), \quad n=0,1,2,\ldots,\\
  &W(x\,;g)=g\log\sin x,\quad g\geq 1/2,\quad 0<x<\pi,\\
  &\mathcal{A}(g)=-\frac{d}{dx}+g\cot x,\quad
  \mathcal{A}(g)^{\dagger}=\frac{d}{dx}+g\cot x,\\
  &\mathcal{H}=-\frac{d^2}{dx^2}+\frac{g(g-1)}{\sin^2x}-g^2.
  \label{oldsutham}
\end{align}
The interpolating Hamiltonian reads simply
\begin{equation}
  \mathcal{H}_s=-\frac{d^2}{dx^2}+\frac{g(g+2s-1)}{\sin^2x}-g^2.
\end{equation}
It is obvious that the interpolating Hamiltonian $\mathcal{H}_s$
has the same  form as the old \eqref{oldsutham} with the coupling
constant $g$ replaced by $g'$ as in the previous case
\eqref{gprimeeq} together with the shift of the ground state
energy $\Delta\mathcal{E}(g,s)=g^{\prime\,2}-g^2$.

%%%%%%%%%%%%%%%%%%%%%%%%%%%%%%%%%%%%%%%%%%%%%%%%%
%                                               %
%  3.4. soliton potential or the symmetric      %
%       Rosen-Morse potential                   %
%                                               %
%%%%%%%%%%%%%%%%%%%%%%%%%%%%%%%%%%%%%%%%%%%%%%%%%
\subsection{soliton potential or the symmetric Rosen-Morse potential}
\label{sRM}

As is well-known $-{g(g+1)/{\cosh^2x}}$ potential is {\em
reflectionless} for integer coupling constant $g$, corresponding
to the KdV soliton. It has a finite number $1+[g]'$ (the greatest
integer not equal or exceeding $g$) of bound states:
\begin{align}
  &\bm{\lambda}=g,\quad \bm{\delta}=-1,\quad
  \mathcal{E}_n(g)=n(2g-n), \quad n=0,1,\ldots,[g]',\\
  &W(x\,;g)=-g\log\cosh x,\quad g>0,\quad -\infty<x<\infty,\\
  &\mathcal{A}(g)=-\frac{d}{dx}-g\tanh x,\quad
  \mathcal{A}(g)^{\dagger}=\frac{d}{dx}-g\tanh x,\\
  &\mathcal{H}=-\frac{d^2}{dx^2}-\frac{g(g+1)}{\cosh^2x}+g^2.
  \label{oldsolham}
\end{align}
The one-parameter Hamiltonian reads simply
\begin{equation}
  \mathcal{H}_s=-\frac{d^2}{dx^2}-\frac{g(g-2s+1)}{\cosh^2x}+g^2.
\end{equation}
It is obvious that the new Hamiltonian $\mathcal{H}_s$ has the
same form as the old \eqref{oldsolham} with the coupling constant
$g$ replaced by $g''$ (for $g>2s-1$ case):
\begin{equation}
  g''=\bigl(-1+\sqrt{1+4g(g+1-2s)}\,\bigr)/2
  \label{secondgprime}
\end{equation}
together with the shift of the ground state energy
$\Delta\mathcal{E}(g,s)=g^2-{g''}^2$.

%%%%%%%%%%%%%%%%%%%%%%%%%%%%%%%%%%%%%%%%%%%%%%%%%
%                                               %
%  3.5. Morse potential                         %
%                                               %
%%%%%%%%%%%%%%%%%%%%%%%%%%%%%%%%%%%%%%%%%%%%%%%%%
\subsection{Morse potential}
\label{Morse}

The prepotential, the Hamiltonian and other data of the Morse
potential read
\begin{align}
  &\bm{\lambda}=(g,\mu),\quad \bm{\delta}=(-1,0),\quad
  \mathcal{E}_n(g,\mu)=n(2g-n), \quad n=0,1,\ldots,[g]',\\
  &W(x\,;g,\mu)=gx-\mu e^x,\quad g,\mu>0,\quad -\infty<x<\infty,\\
  &\mathcal{A}(g,\mu)=-\frac{d}{dx}+g-\mu e^x,\quad
  \mathcal{A}(g,\mu)^{\dagger}=\frac{d}{dx}+g-\mu e^x,\\
  &\mathcal{H}=-\frac{d^2}{dx^2}+\mu^2e^{2x}-\mu(2g+1)e^x+g^2.
\end{align}
The interpolating Hamiltonian reads simply
\begin{equation}
  \mathcal{H}_s=-\frac{d^2}{dx^2}+\mu^2e^{2x}-\mu(2g-2s+1)e^x+g^2.
\end{equation}
This has the same form as the old one with the coupling constant
$g$ replaced by $g'$ (for $g>s$ case):
\begin{equation}
  g'=g-s.
\end{equation}
The shift of the ground state energy is
$\Delta\mathcal{E}(g,\mu,s)=g^2-g^{\prime\,2}$.

%%%%%%%%%%%%%%%%%%%%%%%%%%%%%%%%%%%%%%%%%%%%%%%%%
%                                               %
%  3.6. hyperbolic symmetric top                %
%                                               %
%%%%%%%%%%%%%%%%%%%%%%%%%%%%%%%%%%%%%%%%%%%%%%%%%
\subsection{hyperbolic symmetric top}
\label{hypsymtop}

This system is obtained by the interchange
$\sinh\leftrightarrow\cosh$ from the hyperbolic analog of the
symmetric top problem:
\begin{align}
  &\bm{\lambda}=(g,\mu),\quad \bm{\delta}=(-1,0),\quad
  \mathcal{E}_n(g,\mu)=n(2g-n), \quad n=0,1,\ldots,[g]',\\
  &W(x\,;g,\mu)=-g\log\cosh x-\mu\arctan\sinh x,\quad
  g,\mu>0,\quad -\infty<x<\infty,\\
  &\mathcal{A}(g,\mu)=-\frac{d}{dx}-g\tanh x-\frac{\mu}{\cosh x},\quad
  \mathcal{A}(g,\mu)^{\dagger}=\frac{d}{dx}-g\tanh x-\frac{\mu}{\cosh x},\\
  &\mathcal{H}=-\frac{d^2}{dx^2}
  +\frac{\mu^2-g(g+1)+\mu(2g+1)\sinh x}{\cosh^2 x}+g^2.
\end{align}
The interpolating Hamiltonian reads simply
\begin{equation}
  \mathcal{H}_s=-\frac{d^2}{dx^2}
  +\frac{\mu^2-g(g-2s+1)+\mu(2g-2s+1)\sinh x}{\cosh^2 x}+g^2.
\end{equation}
This has the same form as the old one with the coupling constants
$(g,\mu)$ replaced by $(g',\mu')$, which are determined by
\begin{equation}
  \mu^{\prime\,2}-g'(g'+1)=\mu^2-g(g-2s+1),\quad
  \mu'(2g'+1)=\mu(2g-2s+1).
  \label{eq_for_hst}
\end{equation}
The shift of the ground state energy is
$\Delta\mathcal{E}(g,\mu,s)=g^2-g^{\prime\,2}$. For the boundary
values $s=0$ and $s=1$, $(g',\mu')=(g-s,\mu)$ give a solution of
\eqref{eq_for_hst}.

%%%%%%%%%%%%%%%%%%%%%%%%%%%%%%%%%%%%%%%%%%%%%%%%%
%                                               %
%  3.7. other examples                          %
%                                               %
%%%%%%%%%%%%%%%%%%%%%%%%%%%%%%%%%%%%%%%%%%%%%%%%%
\subsection{other examples}

There are several more examples of shape-invariant quantum
mechanics \cite{susyqm}. But the situation is the same as either
the symmetric P\"oschl-Teller or the symmetric Rosen-Morse
potential case, with the change of the coupling constant
\eqref{gprimeeq}, \eqref{secondgprime}. We present some of them
briefly.

%%%%%%%%%%%%%%%%%%%%%%%%%%%%%%%%%%%%%%%%%%%%%%%%%
%                                               %
%  3.7.1 P\"oschl-Teller potential              %
%                                               %
%%%%%%%%%%%%%%%%%%%%%%%%%%%%%%%%%%%%%%%%%%%%%%%%%
\subsubsection{P\"oschl-Teller potential}
\label{PT}

This is the one-body case of the Sutherland model of $BC$ type:
\begin{align}
  &\bm{\lambda}=(g,h),\quad \bm{\delta}=(1,1),\quad
  \mathcal{E}_n(g,h)=4n(n+g+h), \quad n=0,1,2,\ldots,\\
  &W(x\,;g,h)=g\log\sin x+h\log\cos x,\quad g,h\geq 1/2,\quad 0<x<\pi/2,\\
  &\mathcal{H}=-\frac{d^2}{dx^2}+\frac{g(g-1)}{\sin^2x}
  +\frac{h(h-1)}{\cos^2x}-(g+h)^2,\\
  &\mathcal{H}_s=-\frac{d^2}{dx^2}+\frac{g(g+2s-1)}{\sin^2x}
  +\frac{h(h+2s-1)}{\cos^2x}-(g+h)^2.
\end{align}
The coupling constant $g$ is replaced by $g'$ as \eqref{gprimeeq}
and $h$ by $h'$ similarly. The shift of the ground state energy is
$\Delta\mathcal{E}(g,h,s)=(g'+h')^2-(g+h)^2$.

%%%%%%%%%%%%%%%%%%%%%%%%%%%%%%%%%%%%%%%%%%%%%%%%%
%                                               %
%  3.7.2 Rosen-Morse potential                  %
%                                               %
%%%%%%%%%%%%%%%%%%%%%%%%%%%%%%%%%%%%%%%%%%%%%%%%%
\subsubsection{Rosen-Morse potential}
\label{RM}

The prepotential, the Hamiltonian, etc. are:
\begin{align}
  &\bm{\lambda}=(g,\mu),\quad \bm{\delta}=(-1,0),\\
  &\mathcal{E}_n(g,\mu)=\frac{\mu^2}{g^2}+g^2
  -\Bigl(\frac{\mu^2}{(g-n)^2}+(g-n)^2\Bigr),\quad
  n=0,1,\ldots,\bigl[g-\sqrt{|\mu|}\,\bigr]',\\
  &W(x\,;g,\mu)=-\frac{\mu}{g}x-g\log\cosh x,\quad g>0,\quad -g^2<\mu<g^2,
  \quad -\infty<x<\infty,\\
  &\mathcal{H}=-\frac{d^2}{dx^2}+2\mu\tanh x-\frac{g(g+1)}{\cosh^2x}
  +\frac{\mu^2}{g^2}+g^2,\\
  &\mathcal{H}_s=-\frac{d^2}{dx^2}+2\mu\tanh x-\frac{g(g-2s+1)}{\cosh^2x}
  +\frac{\mu^2}{g^2}+g^2.
\end{align}
The coupling constant $g$ is replaced by $g''$ as
\eqref{secondgprime} (for $g>2s-1$ case) but $\mu$ is unchanged.
The shift of the ground state energy is
$\Delta\mathcal{E}(g,\mu,s)
=\mu^2/g^2+g^2-(\mu^2/g^{\prime\prime\,2}+g^{\prime\prime\,2})$.

%%%%%%%%%%%%%%%%%%%%%%%%%%%%%%%%%%%%%%%%%%%%%%%%%
%                                               %
%  3.7.3 Coulomb potential                      %
%        with the centrifugal barrier           %
%                                               %
%%%%%%%%%%%%%%%%%%%%%%%%%%%%%%%%%%%%%%%%%%%%%%%%%
\subsubsection{Coulomb potential with the centrifugal barrier}
\label{Coulomb}

The prepotential, the Hamiltonian, etc. are:
\begin{align}
  &\bm{\lambda}=(g,\mu),\quad \bm{\delta}=(1,0),\quad
  \mathcal{E}_n(g,\mu)=\frac{\mu^2}{g^2}-\frac{\mu^2}{(g+n)^2},\quad
  n=0,1,2,\ldots,\\
  &W(x\,;g,\mu)=-\frac{\mu}{g}x+g\log x,\quad g\geq\frac12,\quad \mu>0,
  \quad x>0,\\
  &\mathcal{H}=-\frac{d^2}{dx^2}-\frac{2\mu}{x}+\frac{g(g-1)}{x^2}
  +\frac{\mu^2}{g^2},\\
  &\mathcal{H}_s=-\frac{d^2}{dx^2}-\frac{2\mu}{x}+\frac{g(g+2s-1)}{x^2}
  +\frac{\mu^2}{g^2}.
\end{align}
The coupling constant $g$ is replaced by $g'$ as \eqref{gprimeeq}
but $\mu$ is unchanged. The shift of the ground state energy is
$\Delta\mathcal{E}(g,\mu,s)=\mu^2/g^2-\mu^2/g^{\prime\,2}$.

%%%%%%%%%%%%%%%%%%%%%%%%%%%%%%%%%%%%%%%%%%%%%%%%%%%%%%%%%%%%%%%
%                                                             %
%  4. Discrete Quantum Mechanics                              %
%                                                             %
%%%%%%%%%%%%%%%%%%%%%%%%%%%%%%%%%%%%%%%%%%%%%%%%%%%%%%%%%%%%%%%
\section{Discrete Quantum Mechanics}
\label{discreteqm} \setcounter{equation}{0}

The corresponding result for discrete quantum mechanics takes a
slightly different form. Let us start with the brief introduction
of the general setting of the discrete quantum mechanics. For more
details we refer to \cite{os4os5,os6,os7}.

The Hamiltonian of discrete quantum mechanics has a generic form
\begin{equation}
  \mathcal{H}=\sqrt{V(x\,;{\bm\lambda})}\,e^{-i\partial_x}
  \sqrt{V(x\,;{\bm\lambda})^*}
  +\sqrt{V(x\,;{\bm\lambda})^*}\,e^{i\partial_x}\sqrt{V(x\,;{\bm\lambda})}
  -V(x\,;{\bm\lambda})-V(x\,;{\bm\lambda})^*,
  \label{H}
\end{equation}
for various potential functions $V(x\,;{\bm\lambda})$, which are
in general complex. The corresponding Schr\"odinger equation is a
difference equation in stead of a differential equation. It is
also factorised, $\mathcal{H}=\mathcal{A}^{\dagger}\mathcal{A}$,
with
\begin{align}
  \mathcal{A}&=\mathcal{A}({\bm\lambda})
  \eqdef -i\Bigl(e^{-\frac{i}{2}\partial_x}\sqrt{V(x\,;{\bm\lambda})^*}
  -e^{\frac{i}{2}\partial_x}\sqrt{V(x\,;{\bm\lambda})}\Bigr),\\
  \mathcal{A}^{\dagger}&=\mathcal{A}({\bm\lambda})^{\dagger}
  =i\Bigl(\sqrt{V(x\,;{\bm\lambda})}\,e^{-\frac{i}{2}\partial_x}
  -\sqrt{V(x\,;{\bm\lambda})^*}\,e^{\frac{i}{2}\partial_x}\Bigr).
\end{align}
As in the ordinary quantum mechanics cases, the ground state
wavefunction $\phi_0$ is annihilated by the operator
$\mathcal{A}$:
\begin{equation}
  \mathcal{A}(\bm{\lambda})\phi_0(x\,;\bm{\lambda})=0\quad
  \bigl(\Rightarrow\mathcal{H}\phi_0(x\,;\bm{\lambda})=0,\quad
  \mathcal{E}_0(\bm{\lambda})=0\bigr).
  \label{phi0form}
\end{equation}
We introduce a similarity transformed Hamiltonian
$\tilde{\mathcal{H}}(\bm{\lambda})$ in terms of the ground state
wavefunction $\phi_0=\phi_0(x\,;\bm{\lambda})$:
\begin{equation}
  \tilde{\mathcal{H}}\eqdef
  \phi_0^{-1}\circ\mathcal{H}\circ\phi_0
  =V(x\,;{\bm\lambda})e^{-i\partial_x}+V(x\,;{\bm\lambda})^*e^{i\partial_x}
  -V(x\,;{\bm\lambda})-V(x\,;{\bm\lambda})^*.
  \label{tilH}
\end{equation}
It acts on the polynomial part of the eigenfunction
\begin{align}
  &\phi_n(x\,;\bm{\lambda})\propto
  \phi_0(x\,;\bm{\lambda})P_n(\eta(x)\,;\bm{\lambda}),
  \label{phinphi0Pn}\\
  &\tilde{\mathcal{H}}P_n(\eta(x)\,;\bm{\lambda})
  =\mathcal{E}_n(\bm{\lambda})P_n(\eta(x)\,;\bm{\lambda}).
\end{align}
Here $P_n(\eta\,;\bm{\lambda})$ is a polynomial of degree $n$ in
variable $\eta$, and $\eta(x)$ is a real function of $x$. For all
the examples given in this section, the eigenfunctions have this
form.

We assume that the system is  shape-invariant \eqref{shapeinv}.
Next let us introduce the similarity transformation of the
reversed order Hamiltonian
$\mathcal{H}_r(\bm{\lambda})=\mathcal{A}(\bm{\lambda})
\mathcal{A}(\bm{\lambda})^{\dagger}$:
\begin{align}
  &\tilde{\mathcal{H}}_r\eqdef
  \phi_{r0}^{-1}\circ\mathcal{H}_r\circ\phi_{r0}\n
  &=V(x\,;\bm{\lambda}+\bm{\delta})e^{-i\partial_x}
  +V(x\,;\bm{\lambda}+\bm{\delta})^*e^{i\partial_x}
  -V(x\,;\bm{\lambda}+\bm{\delta})-V(x\,;\bm{\lambda}+\bm{\delta})^*
  +\mathcal{E}_1(\bm{\lambda}),
  \label{tilH2}
\end{align}
with respect to its ground state wavefunction
$\phi_{r0}=\phi_{r0}(x\,;\bm{\lambda})$ defined by
\begin{equation}
  \mathcal{A}(\bm{\lambda}+\bm{\delta})\phi_{r0}(x\,;\bm{\lambda})=0\quad
  \bigl(\Rightarrow \phi_{r0}(x\,;\bm{\lambda})\propto
  \phi_0(x\,;\bm{\lambda}+\bm{\delta})\bigr).
\end{equation}
We introduce the one parameter family of (similarity transformed)
Hamiltonians:
\begin{align}
  \tilde{\mathcal{H}}_s&\eqdef
  (1-s)\tilde{\mathcal{H}}+s\tilde{\mathcal{H}}_r
  \label{tHs}\\
  &=V_s(x\,;\bm{\lambda})e^{-i\partial_x}
  +V_s(x\,;\bm{\lambda})^*e^{i\partial_x}
  -V_s(x\,;\bm{\lambda})-V_s(x\,;\bm{\lambda})^*
  +s\mathcal{E}_1(\bm{\lambda}),
\end{align}
which depends on the interpolated potential function
$V_s(x\,;\bm{\lambda})$:
\begin{equation}
  V_s(x\,;\bm{\lambda})\eqdef
  (1-s)V(x\,;\bm{\lambda})+s\,V(x\,;\bm{\lambda}+\bm{\delta}).
\end{equation}
Since both $\tilde{\mathcal{H}}$ and $\tilde{\mathcal{H}}_r$ act
on the space of polynomials, their sum is meaningful and acts on
the same space.

Our assertion is that $V_s(x\,;\bm{\lambda})$ is simply the same
potential function with shifted coupling constants
$\bm{\lambda}'$:
\begin{equation}
  V_s(x\,;\bm{\lambda})=V(x\,;\bm{\lambda}').
\end{equation}
Then we have
\begin{equation}
  \tilde{\mathcal{H}}_s=\tilde{\mathcal{H}}(\bm{\lambda}')
  +\widetilde{\Delta\mathcal{E}}(\bm{\lambda},s),\quad
  \widetilde{\Delta\mathcal{E}}(\bm{\lambda},s)=s\mathcal{E}_1(\bm{\lambda}).
\end{equation}
Recall that $\bm{\lambda}'$ depend on $s$. The eigenfunctions and
eigenvalues of $\tilde{\mathcal{H}}_s$ are given by
\begin{equation}
  \tilde{\mathcal{H}}_sP_n(\eta(x)\,;\bm{\lambda}')
  =\bigl(\mathcal{E}_n(\bm{\lambda}')+s\mathcal{E}_1(\bm{\lambda})\bigr)
  P_n(\eta(x)\,;\bm{\lambda}'),\quad n=0,1,2,\ldots.
  \label{eigen_tHs}
\end{equation}
By similarity transformation inversely with respect to
$\phi_{s0}=\phi_{s0}(x\,;\bm{\lambda})$ defined by
\[
\mathcal{A}(\bm{\lambda}')\phi_{s0}(x\,;\bm{\lambda})=0 \quad
\Longrightarrow
\phi_{s0}(x\,;\bm{\lambda})\propto\phi_0(x\,;\bm{\lambda}'),
\]
we introduce $\check{\mathcal{H}}_s$ as
\begin{align}
  \check{\mathcal{H}}_s&\eqdef
  \phi_{s0}\circ\tilde{\mathcal{H}}_s\circ\phi_{s0}^{-1}
  \label{checkHs}\\
  &=(1-s)\frac{\phi_0(x\,;\bm{\lambda}')}{\phi_0(x\,;\bm{\lambda})}
  \circ\mathcal{H}\circ
  \frac{\phi_0(x\,;\bm{\lambda})}{\phi_0(x\,;\bm{\lambda}')}
  +s\frac{\phi_0(x\,;\bm{\lambda}')}{\phi_0(x\,;\bm{\lambda}+\bm{\delta})}
  \circ\mathcal{H}_r\circ
  \frac{\phi_0(x\,;\bm{\lambda}+\bm{\delta})}{\phi_0(x\,;\bm{\lambda}')}\n
  &=\mathcal{H}(\bm{\lambda}')+s\mathcal{E}_1(\bm{\lambda}).
\end{align}
This one parameter family of Hamiltonians $\check{\mathcal{H}}_s$
interpolates $\mathcal{H}$ ($s=0$) and $\mathcal{H}_r$ ($s=1$).
Its eigenfunctions and eigenvalues are given by
\begin{equation}
  \check{\mathcal{H}}_s\phi_n(x\,;\bm{\lambda}')
  =\bigl(\mathcal{E}_n(\bm{\lambda}')+s\mathcal{E}_1(\bm{\lambda})\bigr)
  \phi_n(x\,;\bm{\lambda}'),\quad n=0,1,2,\ldots.
  \label{eigen_cHs}
\end{equation}

{}From this our claim follows
\begin{quote}
  the interpolated (similarity transformed) Hamiltonian
  $\check{\mathcal{H}}_s$ ($\tilde{\mathcal{H}}_s$) ($0\le s\le 1$)
  also describes the same {\em shape-invariant} and
  {\em exactly solvable} system with shifted {\em coupling constants}.
\end{quote}
This claim can be easily verified for each known shape-invariant
discrete quantum mechanics given below. We will denote them by the
name of the polynomials $\{P_n(\eta)\}$ constituting the
eigenfunctions.

%%%%%%%%%%%%%%%%%%%%%%%%%%%%%%%%%%%%%%%%%%%%%%%%%
%                                               %
%  4.1. Meixner-Pollaczek polynomial            %
%                                               %
%%%%%%%%%%%%%%%%%%%%%%%%%%%%%%%%%%%%%%%%%%%%%%%%%
\subsection{Meixner-Pollaczek polynomial}

The Meixner-Pollaczek polynomial \cite{os4os5,koeswart,degruij} is
a one-parameter deformation of the Hermite polynomial. Therefore
the corresponding discrete quantum mechanics is also called a {\em
deformed harmonic oscillator\/}. Its potential function and other
data are:
\begin{align}
  &\bm{\lambda}=\lambda,\quad \bm{\delta}=1/2,\quad
  \mathcal{E}_n(\lambda)=2n, \quad n=0,1,2,\ldots,\\
  &V(x\,;\lambda)=\lambda+i x,\quad \lambda>0,\quad -\infty<x<\infty.
\end{align}
The increase of the ground state energy
$\mathcal{E}_1(\bm{\lambda})=2$ is independent of the coupling
constant as for the ordinary harmonic oscillator. It is rather
trivial to verify our assertion:
\begin{equation}
  V_s(x\,;\lambda)=\lambda+s/2+i x,
\end{equation}
and
\begin{equation}
  \lambda'=\lambda+s/2,\quad \widetilde{\Delta\mathcal{E}}(\lambda\,;s)=2s.
\end{equation}

%%%%%%%%%%%%%%%%%%%%%%%%%%%%%%%%%%%%%%%%%%%%%%%%%
%                                               %
%  4.2. continuous Hahn polynomial              %
%                                               %
%%%%%%%%%%%%%%%%%%%%%%%%%%%%%%%%%%%%%%%%%%%%%%%%%
\subsection{continuous Hahn polynomial}

The continuous Hahn polynomial with special parameters
\cite{os4os5,koeswart} is a two-parameter deformation of the
Hermite polynomial. The potential function and the other data are:
\begin{align}
  &\bm{\lambda}=(a,b),\quad \bm{\delta}=(1/2,1/2),\quad
  \mathcal{E}_n(\bm{\lambda})=n(n+2a+2b-1), \quad n=0,1,2,\ldots,\\
  &V(x\,;\bm{\lambda})=(a+ix)(b+ix),\quad a,b>0,\quad -\infty<x<\infty.
\end{align}
These lead to $\bm{\lambda}'=(a',b')$ determined by
\begin{equation}
  a'+b'=a+b+s,\quad a'b'=ab+(a+b)s/2+s/4
  \label{eq_for_cHahn}
\end{equation}
and
\begin{equation}
  \widetilde{\Delta\mathcal{E}}(\bm{\lambda}\,;s)=2(a+b)s.
\end{equation}
For the boundary values $s=0$ and $s=1$, $(a',b')=(a+s/2,b+s/2)$
is a solution of \eqref{eq_for_cHahn}.

%%%%%%%%%%%%%%%%%%%%%%%%%%%%%%%%%%%%%%%%%%%%%%%%%
%                                               %
%  4.3. continuous dual Hahn polynomial         %
%                                               %
%%%%%%%%%%%%%%%%%%%%%%%%%%%%%%%%%%%%%%%%%%%%%%%%%
\subsection{continuous dual Hahn polynomial}

The continuous dual Hahn polynomial \cite{os4os5,koeswart} is a
two-parameter deformation of the Laguerre polynomial. Therefore
the corresponding quantum mechanics is a deformation of the
potential $x^2+1/x^2$, or the one body case of the Calogero model
\cite{cal} discussed in \S \ref{ho_cbp}. The potential function
and the other data are:
\begin{align}
  &\bm{\lambda}=(a,b,c),\quad \bm{\delta}=(1/2,1/2,1/2),\quad
  \mathcal{E}_n(\bm{\lambda})=n, \quad n=0,1,2,\ldots,\\
  &V(x\,;\bm{\lambda})=\frac{(a+ix)(b+ix)(c+ix)}{2ix(2ix+1)},\quad
  a,b,c>0,\quad 0<x<\infty.
   \label{contdualhahn}
\end{align}
As in the Calogero case, the increase of the ground state energy
is independent of the coupling constants and this leads to the
linear energy spectrum. The shifted parameters
$\bm{\lambda}'=(a',b',c')$ are determined by
\begin{align}
  a'+b'+c'&=a+b+c+3s/2,
  \label{eq_for_cdHahn_1}\\
  a'b'+b'c'+c'a'&=ab+bc+ca+(a+b+c)s+3s/4,\\
  a'b'c'&=abc+(ab+bc+ca)s/2+(a+b+c)s/4+s/8
  \label{eq_for_cdHahn_3}
\end{align}
and
\begin{equation}
  \widetilde{\Delta\mathcal{E}}(\bm{\lambda}\,;s)=s.
\end{equation}
For the boundary values $s=0$ and $s=1$,
$(a',b',c')=(a+s/2,b+s/2,c+s/2)$ is a solution of
\eqref{eq_for_cdHahn_1}--\eqref{eq_for_cdHahn_3}.

%%%%%%%%%%%%%%%%%%%%%%%%%%%%%%%%%%%%%%%%%%%%%%%%%
%                                               %
%  4.4. Wilson polynomial                       %
%                                               %
%%%%%%%%%%%%%%%%%%%%%%%%%%%%%%%%%%%%%%%%%%%%%%%%%
\subsection{Wilson polynomial}

The Wilson polynomial \cite{os4os5,koeswart} is a three-parameter
deformation of the Laguerre polynomial. The potential function and
the other data are:
\begin{align}
  &\bm{\lambda}=(a,b,c,d),\quad
  \bm{\delta}=(1/2,1/2,1/2,1/2),\\
  &\mathcal{E}_n(\bm{\lambda})=n(n+a+b+c+d-1), \qquad \qquad \quad
n=0,1,2,\ldots,\\
  &V(x\,;\bm{\lambda})=\frac{(a+ix)(b+ix)(c+ix)(d+ix)}{2ix(2ix+1)},\quad
  a,b,c,d>0,\quad 0<x<\infty.
  \label{Wilsonpoly}
\end{align}
The shifted parameters $\bm{\lambda}'=(a',b',c',d')$ are
determined by
\begin{align}
  &a'+b'+c'+d'=a+b+c+d+2s,
  \label{eq_for_Wilson_1}\\
  &a'b'+a'c'+a'd'+b'c'+b'd'+c'd'\n
  &\quad=ab+ac+ad+bc+bd+cd+3(a+b+c+d)s/2+3s/2,\\
  &a'b'c'+a'b'd'+a'c'd'+b'c'd'\\
  &\quad=abc+abd+acd+bcd
  +(ab+ac+ad+bc+bd+cd)s+3(a+b+c+d)s/4+s/2,\n
  &a'b'c'd'=abcd+(abc+abd+acd+bcd)s/2+(ab+ac+ad+bc+bd+cd)s/4\n
  &\qquad\qquad\qquad
  +(a+b+c+d)s/8+s/16
  \label{eq_for_Wilson_4}
\end{align}
and
\begin{equation}
  \widetilde{\Delta\mathcal{E}}(\bm{\lambda}\,;s)=(a+b+c+d)s.
\end{equation}
For the boundary values $s=0$ and $s=1$,
$(a',b',c',d')=(a+s/2,b+s/2,c+s/2,d+s/2)$ is a solution of
\eqref{eq_for_Wilson_1}--\eqref{eq_for_Wilson_4}.

%%%%%%%%%%%%%%%%%%%%%%%%%%%%%%%%%%%%%%%%%%%%%%%%%
%                                               %
%  4.5. Askey-Wilson polynomial                 %
%                                               %
%%%%%%%%%%%%%%%%%%%%%%%%%%%%%%%%%%%%%%%%%%%%%%%%%
\subsection{Askey-Wilson polynomial}

The Askey-Wilson polynomial \cite{os4os5,koeswart} belongs to a
different type of shape-invariant discrete quantum mechanical
single particle systems. It has a multiplicative shift of
parameters rather than the additive shift \eqref{shapeinv}
discussed in the rest of this paper.

Following \cite{os4os5,os6,os7}, we use variables $\theta$, $x$
and $z$, which are related as
\begin{equation}
  0\leq\theta\leq\pi,\quad x=\cos\theta,\quad z=e^{i\theta}.
\end{equation}
The dynamical variable is $\theta$ and the inner product is
$\langle f|g\rangle=\int_0^{\pi}d\theta f(\theta)^*g(\theta)$. We
denote $D\eqdef z\frac{d}{dz}$. Then $q^D$ is a $q$-shift
operator, $q^Df(z)=f(qz)$. We assume $0<q<1$.  We note here that
\begin{equation}
  \int_0^{\pi}d\theta=\int_{-1}^1\frac{dx}{\sqrt{1-x^2}},\quad
  -i\frac{d}{d\theta} =z\frac{d}{dz}=D.
\end{equation}
The Hamiltonian is
\begin{equation}
  \mathcal{H}=\sqrt{V(z\,;\bm{\lambda})}\,q^{D}\!\sqrt{V(z\,;\bm{\lambda})^*}
  +\sqrt{V(z\,;\bm{\lambda})^*}\,q^{-D}\!\sqrt{V(z\,;\bm{\lambda})}
  -V(z\,;\bm{\lambda})-V(z\,;\bm{\lambda})^*,
  \label{H-q}
\end{equation}
which is factorized, i.e.
$\mathcal{H}=\mathcal{A}^{\dagger}\mathcal{A}$, with
\begin{align}
  \mathcal{A}&=\mathcal{A}(\bm{\lambda})\eqdef
  i\Bigl(q^{\frac{D}{2}}\sqrt{V(z\,;\bm{\lambda})^*}
  -q^{-\frac{D}{2}}\sqrt{V(z\,;\bm{\lambda})}\Bigr),\\
  \mathcal{A}^{\dagger}&=\mathcal{A}(\bm{\lambda})^{\dagger}=
  -i\Bigl(\sqrt{V(z\,;\bm{\lambda})}\,q^{\frac{D}{2}}
  -\sqrt{V(z\,;\bm{\lambda})^*}\,q^{-\frac{D}{2}}\Bigr).
\end{align}
The ground state $\phi_0$ is the function annihilated by
$\mathcal{A}$:
\begin{equation}
  \mathcal{A}(\bm{\lambda})\phi_0(z\,;\bm{\lambda})=0\quad
  (\Rightarrow \mathcal{H}\phi_0(z\,;\bm{\lambda})=0,\quad
  \mathcal{E}_0(\bm{\lambda})=0).
  \label{ground}
\end{equation}
The potential function and other data are
\begin{align}
  &\bm{\lambda}=(a,b,c,d),\quad -1<a,b,c,d<1,\quad abcd<q,\\
  &\mathcal{E}_n(\bm{\lambda})=(q^{-n}-1)(1-abcdq^{n-1}),\qquad
  n=0,1,2,\ldots,\\
  &V(z\,;\bm{\lambda})=\frac{(1-az)(1-bz)(1-cz)(1-dz)}{(1-z^2)(1-qz^2)}.
\end{align}
The similarity transformed Hamiltonian in terms of the ground
state wavefunction $\phi_0=\phi_0(z\,;\bm{\lambda})$ reads
\begin{equation}
  \tilde{\mathcal{H}}\eqdef
  \phi_0^{-1}\circ \mathcal{H}\circ\phi_0
  =V(z\,;\bm{\lambda})\,q^{D}+V(z\,;\bm{\lambda})^*\,q^{-D}
  -V(z\,;\bm{\lambda})-V(z\,;\bm{\lambda})^*.
  \label{tildeH}
\end{equation}

The shape-invariance relation reads
\begin{equation}
  \mathcal{A}(\bm{\lambda})\mathcal{A}(\bm{\lambda})^{\dagger}
  =q^{-1}\mathcal{A}(q^{\frac12}\bm{\lambda})^{\dagger}
  \mathcal{A}(q^{\frac12}\bm{\lambda})
  +\mathcal{E}_1(\bm{\lambda}).
  \label{AAd=AdA+E}
\end{equation}
The similarity transformed reversed order Hamiltonian
$\mathcal{H}_r(\bm{\lambda})=\mathcal{A}(\bm{\lambda})
\mathcal{A}(\bm{\lambda})^{\dagger}$, with respect to its ground
state wavefunction $\phi_{r0}=\phi_{r0}(z\,;\bm{\lambda})$
($\mathcal{A}(q^{\frac12}\bm{\lambda})\phi_{r0}(z\,;\bm{\lambda})=0$
$\Rightarrow$
$\phi_{r0}(z\,;\bm{\lambda})\propto\phi_0(z\,;q^{\frac12}\bm{\lambda})$),
reads
\begin{align}
  &\tilde{\mathcal{H}}_r\eqdef
  \phi_{r0}^{-1}\circ\mathcal{H}_r\circ\phi_{r0}\n
  &=q^{-1}\Bigl(V(z\,;q^{\frac12}\bm{\lambda})\,q^{D}
  +V(z\,;q^{\frac12}\bm{\lambda})^*\,q^{-D}
  -V(z\,;q^{\frac12}\bm{\lambda})-V(z\,;q^{\frac12}\bm{\lambda})^*\Bigr)
  +\mathcal{E}_1(\bm{\lambda}).
  \label{tilH3}
\end{align}
The one parameter family of (similarity transformed) Hamiltonians
is introduced as
\begin{align}
  \tilde{\mathcal{H}}_s&\eqdef
  (1-s)\tilde{\mathcal{H}}+s\tilde{\mathcal{H}}_r\n
  &=V_s(z\,;\bm{\lambda})\,q^{D}+V_s(z\,;\bm{\lambda})^*\,q^{-D}
  -V_s(z\,;\bm{\lambda})-V_s(z\,;\bm{\lambda})^*
  +s\mathcal{E}_1(\bm{\lambda}),
\end{align}
which depends on the interpolated potential function
$V_s(z\,;\bm{\lambda})$:
\begin{equation}
  V_s(z\,;\bm{\lambda})\eqdef(1-s)V(z\,;\bm{\lambda})
  +sq^{-1}V(z\,;q^{\frac12}\bm{\lambda}).
\end{equation}

Our assertion is that $V_s(z\,;\bm{\lambda})$ is simply the same
potential function with shifted coupling constants $\bm{\lambda}'
=(a',b',c',d')$ with an overall scaling factor $\alpha>0$:
\begin{equation}
  V_s(z\,;\bm{\lambda})=\alpha V(z\,;\bm{\lambda}').
\end{equation}
Then we have
\begin{equation}
  \tilde{\mathcal{H}}_s=\alpha\tilde{\mathcal{H}}(\bm{\lambda}')
  +\widetilde{\Delta\mathcal{E}}(\bm{\lambda},s),\quad
  \widetilde{\Delta\mathcal{E}}(\bm{\lambda},s)=s\mathcal{E}_1(\bm{\lambda}).
\end{equation}
We can define $\check{\mathcal{H}}_s$ like as \eqref{checkHs}.
Equations like as \eqref{eigen_tHs} and \eqref{eigen_cHs} hold
also.

The shifted parameters are determined by
\begin{align}
  \alpha&=1+(q^{-1}-1)s,
  \label{eq_for_AW_1}\\
  a'+b'+c'+d'&=\frac{1+(q^{-1/2}-1)s}{1+(q^{-1}-1)s}(a+b+c+d),\\
  a'b'+a'c'+a'd'+b'c'+b'd'+c'd'
  &=\frac{1}{1+(q^{-1}-1)s}(ab+ac+ad+bc+bd+cd),\\
  a'b'c'+a'b'd'+a'c'd'+b'c'd'
  &=\frac{1-(1-q^{1/2})s}{1+(q^{-1}-1)s}(abc+abd+acd+bcd),\\
  a'b'c'd'&=\frac{1-(1-q)s}{1+(q^{-1}-1)s}\,abcd
  \label{eq_for_AW_5}
\end{align}
and
\begin{equation}
  \widetilde{\Delta\mathcal{E}}(\bm{\lambda},s)=(q^{-1}-1)(1-abcd)s.
\end{equation}
For the boundary values $s=0$ and $s=1$, $\alpha=q^{-s}$ and
$(a',b',c',d')=q^{s/2}(a,b,c,d)$ is a solution of
\eqref{eq_for_AW_1}--\eqref{eq_for_AW_5}.

%%%%%%%%%%%%%%%%%%%%%%%%%%%%%%%%%%%%%%%%%%%%%%%%%%%%%%%%%%%%%%%
%                                                             %
%  5. Ordinary Quantum Mechanics Revisited                    %
%                                                             %
%%%%%%%%%%%%%%%%%%%%%%%%%%%%%%%%%%%%%%%%%%%%%%%%%%%%%%%%%%%%%%%
\section{Ordinary Quantum Mechanics Revisited}
\label{qmrevisited} \setcounter{equation}{0}

In the previous section we have considered the one parameter
family of (similarity transformed) Hamiltonians \eqref{tHs},
\eqref{checkHs}. This construction method applies also to the
ordinary quantum mechanics studied in section \ref{ordinaryqm}.

We introduce a similarity transformed Hamiltonian
$\tilde{\mathcal{H}}$ in terms of the ground state wavefunction
$\phi_0=\phi_0(x\,;\bm{\lambda})\propto e^{W(x\,;\bm{\lambda})}$:
\begin{equation}
  \tilde{\mathcal{H}}\eqdef
  \phi_0^{-1}\circ\mathcal{H}\circ\phi_0
  =-\frac{d^2}{dx^2}-2\frac{dW(x\,;\bm{\lambda})}{dx}\frac{d}{dx}.
\end{equation}
Next let us introduce the similarity transformation of the
reversed order Hamiltonian \eqref{Hr}, \eqref{shapeinv} with
respect to its ground state wavefunction
$\phi_{r0}=\phi_{r0}(x\,;\bm{\lambda})\propto
e^{W(x\,;\bm{\lambda}+\bm{\delta})}$:
\begin{equation}
  \tilde{\mathcal{H}}_r\eqdef
  \phi_{r0}^{-1}\circ\mathcal{H}_r\circ\phi_{r0}
  =-\frac{d^2}{dx^2}
  -2\frac{dW(x\,;\bm{\lambda}+\bm{\delta})}{dx}\frac{d}{dx}
  +\mathcal{E}_1(\bm{\lambda}).
\end{equation}
The one parameter family of similarity transformed Hamiltonians is
introduced as \eqref{tHs}:
\begin{equation}
  \tilde{\mathcal{H}}_s\eqdef
  (1-s)\tilde{\mathcal{H}}+s\tilde{\mathcal{H}}_r
  =-\frac{d^2}{dx^2}
  -2\frac{dW_s(x\,;\bm{\lambda})}{dx}\frac{d}{dx}
  +s\mathcal{E}_1(\bm{\lambda}),
\end{equation}
where the interpolated prepotential $W_s(x\,;\bm{\lambda})$ is
given by
\begin{equation}
  W_s(x\,;\bm{\lambda})\eqdef(1-s)W(x\,;\bm{\lambda})
  +sW(x\,;\bm{\lambda}+\bm{\delta}).
\end{equation}
We do not assume \eqref{phinphi0Pn} in general.

Our assertion is that $W_s(x\,;\bm{\lambda})$ is simply the same
prepotential with shifted coupling constants $\bm{\lambda}'$:
\begin{equation}
  W_s(x\,;\bm{\lambda})=W(x\,;\bm{\lambda}').
\end{equation}
Then we have
\begin{equation}
  \tilde{\mathcal{H}}_s=\tilde{\mathcal{H}}(\bm{\lambda}')
  +\widetilde{\Delta\mathcal{E}}(\bm{\lambda},s),\quad
  \widetilde{\Delta\mathcal{E}}(\bm{\lambda},s)=s\mathcal{E}_1(\bm{\lambda}).
\end{equation}
Note that $\bm{\lambda}'$ depend on $s$. As in \eqref{checkHs}, we
define inversely similarity transformed Hamiltonian with respect
to $\phi_{s0}=\phi_{s0}(x\,;\bm{\lambda}) \propto
e^{W_s(x\,;\bm{\lambda})}=e^{W(x\,;\bm{\lambda}')}$:
\begin{align}
  \check{\mathcal{H}}_s&\eqdef
  \phi_{s0}\circ\tilde{\mathcal{H}}_s\circ\phi_{s0}^{-1}\n
  &=(1-s)\frac{\phi_0(x\,;\bm{\lambda}')}{\phi_0(x\,;\bm{\lambda})}
  \circ\mathcal{H}\circ
  \frac{\phi_0(x\,;\bm{\lambda})}{\phi_0(x\,;\bm{\lambda}')}
  +s\frac{\phi_0(x\,;\bm{\lambda}')}{\phi_0(x\,;\bm{\lambda}+\bm{\delta})}
  \circ\mathcal{H}_r\circ
  \frac{\phi_0(x\,;\bm{\lambda}+\bm{\delta})}{\phi_0(x\,;\bm{\lambda}')}\n
  &=\mathcal{H}(\bm{\lambda}')+s\mathcal{E}_1(\bm{\lambda}).
\end{align}
This one parameter family of Hamiltonians $\check{\mathcal{H}}_s$
interpolates $\mathcal{H}$ ($s=0$) and $\mathcal{H}_r$ ($s=1$),
and its eigenfunctions and eigenvalues have the same form as those
in \eqref{eigen_cHs}. This interpolation is different from the one
studied in section \ref{ordinaryqm}.

For the quantum mechanical systems studied in section
\ref{ordinaryqm}, we give a list of $\bm{\lambda}'$:
\begin{alignat}{2}
  &\S \ref{ho}&\ \ &:\quad
  \omega'=\omega,\\
  &\S \ref{ho_cbp}&\ \ &:\quad
  \omega'=\omega,\quad g'=g+s,\\
  &\S \ref{sPT}&&:\quad
  g'=g+s,\\
  &\S \ref{sRM}&&:\quad
  g'=g-s,\\
  &\S \ref{Morse}&&:\quad
  g'=g-s,\quad \mu'=\mu,\\
  &\S \ref{hypsymtop}&&:\quad
  g'=g-s,\quad \mu'=\mu,\\
  &\S \ref{PT}&&:\quad
  g'=g+s,\quad h'=h+s,\\
  &\S \ref{RM}&&:\quad
  g'=g-s,\quad \mu'=\frac{(g-s)(g-1+s)}{g(g-1)}\,\mu,\\
  &\S \ref{Coulomb}&&:\quad
  g'=g+s,\quad \mu'=\frac{(g+s)(g+1-s)}{g(g+1)}\,\mu.
\end{alignat}

%%%%%%%%%%%%%%%%%%%%%%%%%%%%%%%%%%%%%%%%%%%%%%%%%%%%%%%%%%%%%%%
%                                                             %
%  6. Summary and Comments                                    %
%                                                             %
%%%%%%%%%%%%%%%%%%%%%%%%%%%%%%%%%%%%%%%%%%%%%%%%%%%%%%%%%%%%%%%
\section{Summary and Comments}

Interpolation of the SUSY-partner Hamiltonians are discussed for
various shape-invariant quantum mechanics including the discrete
ones. For ordinary quantum mechanics we present two kinds of
interpolation, ${\mathcal{H}}_s$ and $\check{\mathcal{H}}_s$ (or
$\tilde{\mathcal{H}}_s$). The shape-invariance is inherited by the
interpolating Hamiltonian with shifted coupling constants. The
interpolation works at the corresponding eigenfunctions level. For
the quantum systems whose eigenfunctions have the form
\eqref{phinphi0Pn}, the annihilation-creation operators can be
constructed \cite{os7}. These annihilation-creation operators are
also interpolated. The eigenfunctions of annihilation operators
are interpreted as coherent states. Therefore these coherent
states are also interpolated.

It would be interesting if a similar interpolation idea works for
a wider class of quantum mechanics \cite{morris} which do not have
shape-invariance.

%%%%%%%%%%%%%%%%%%%%%%%%%%%%%%%%%%%%%%%%%%%%%%%%%%%%%%%%%%%%%%%
%                                                             %
%  Acknowledgements                                           %
%                                                             %
%%%%%%%%%%%%%%%%%%%%%%%%%%%%%%%%%%%%%%%%%%%%%%%%%%%%%%%%%%%%%%%
\section*{Acknowledgements}

This work is supported in part by Grants-in-Aid for Scientific
Research from the Ministry of Education, Culture, Sports, Science
and Technology, No.18340061 and No.19540179, in part by the U.S.
National Science Foundation Grant No. PHY-0555231 at the
University of Wisconsin and in part by the University of Wisconsin
Research Committee with funds granted by the Wisconsin Alumni
Research Foundation.

%%%%%%%%%%%%%%%%%%%%%%%%%%%%%%%%%%%%%%%%%%%%%%%%%%%%%%%%%%%%%%%
%                                                             %
%  References                                                 %
%                                                             %
%%%%%%%%%%%%%%%%%%%%%%%%%%%%%%%%%%%%%%%%%%%%%%%%%%%%%%%%%%%%%%%

\end{document}